\documentclass[twocolumn,english,showpacs,preprintnumbers,amsmath,amssymb,aps,prb]{revtex4}
\usepackage[T1]{fontenc}
\usepackage[latin9]{inputenc}
\usepackage{graphicx}
\usepackage{esint}
\usepackage{epstopdf}

\makeatletter

\@ifundefined{textcolor}{}
{%
 \definecolor{BLACK}{gray}{0}
 \definecolor{WHITE}{gray}{1}
 \definecolor{RED}{rgb}{1,0,0}
 \definecolor{GREEN}{rgb}{0,1,0}
 \definecolor{BLUE}{rgb}{0,0,1}
 \definecolor{CYAN}{cmyk}{1,0,0,0}
 \definecolor{MAGENTA}{cmyk}{0,1,0,0}
 \definecolor{YELLOW}{cmyk}{0,0,1,0}
 }

\makeatother

\usepackage{babel}
\begin{document}

\title{Strongly Enhanced Vortex Pinning by Conformal Crystal Arrays}

\author{D. Ray$^{1,2}$, C. J. Olson Reichhardt$^{2}$, B. Jank{ó}$^{1}$,
and C. Reichhardt$^{2}$}

\affiliation{$^{1}$Department of Physics, University of Notre Dame, Notre Dame, Indiana 46556 \\
$^{2}$Theoretical Division, Los Alamos National Laboratory, Los
Alamos, New Mexico 87545 }

\date{\today}
\begin{abstract}
Conformal crystals are non-uniform structures created by 
a conformal 
transformation 
of regular two-dimensional 
lattices.
We show that gradient-driven vortices interacting with a conformal
pinning array exhibit substantially 
stronger pinning effects over a much larger range of field than found
for 
random or periodic pinning arrangements. The pinning
enhancement is partially due to 
matching of the critical
flux gradient with the pinning gradient,
but the preservation of the sixfold ordering in the conformally transformed
hexagonal lattice plays a crucial role.
Our results can be generalized to a wide class of gradient-driven
interacting particle systems such as colloids on optical trap arrays.
\end{abstract}


\maketitle
One of the most important problems for applications of type-II superconductors
is how to create high critical currents or strong vortex pinning over
a wide range of applied magnetic fields \cite{Blatter}.  
For over sixty years, it has been understood that the ground state vortex
structure is a hexagonal lattice
\cite{Alexi}, so many methods have been developed 
to 
increase the critical current using uniform pinning arrays that incorporate
periodicity to match the vortex structure
\cite{Baert,met,Bending,Welp,R,Martin,Schuller,Peeters,block1,block2,block3}.
The pinning is enhanced at commensurate fields when the number of vortices
equals an integer multiple of the number of pinning sites, but away from
these specific matching fields, the enhancement of the critical current is lost
\cite{Periodic}.
Efforts
to enhance the pinning at incommensurate fields have included the
use of quasicrystalline substrates \cite{Kemmler} or diluted periodic
arrays \cite{Reichhardt,Kleiner,Ros,Olson,Thakur}, where studies
show that new types of non-integer commensurate states can arise in
addition to the integer matching configurations. Hyperbolic tessellation
arrays were also recently considered \cite{Mikso}. 

Part of the problem is the fact that under an applied current, the vortex
structure does not remain uniform but instead develops a 
Bean-like flux gradient \cite{Bean}:
the vortex density is highest at the edges of the sample
when the magnetic field is increased, and highest in the center of
the sample when the magnetic field is removed and only trapped flux
remains inside the sample. 
As a consequence, uniform pinning arrays
generally have a portion of the pinning sites that are not fully occupied,
suggesting that a more optimal pinning arrangement should include
some type of density gradient to match the critical flux gradient.
Here we show that a novel type of pinning structure, created using a
conformal 
transformation of a uniform hexagonal lattice, 
produces a much higher critical current over a much wider range of magnetic
fields than any pinning geometry considered up until now.  
Conformal crystals not only have a density gradient, but also
preserve aspects of the hexagonal ordering naturally adopted by the
vortex lattice. 
The pinning
enhancement we find is substantial and will be very important for a wide
range of superconductor applications and flux control.  Our results can
also be important for stabilizing novel self-assembled structures created
using density gradients, such as in colloidal systems \cite{Grier}.

\begin{figure}
\includegraphics{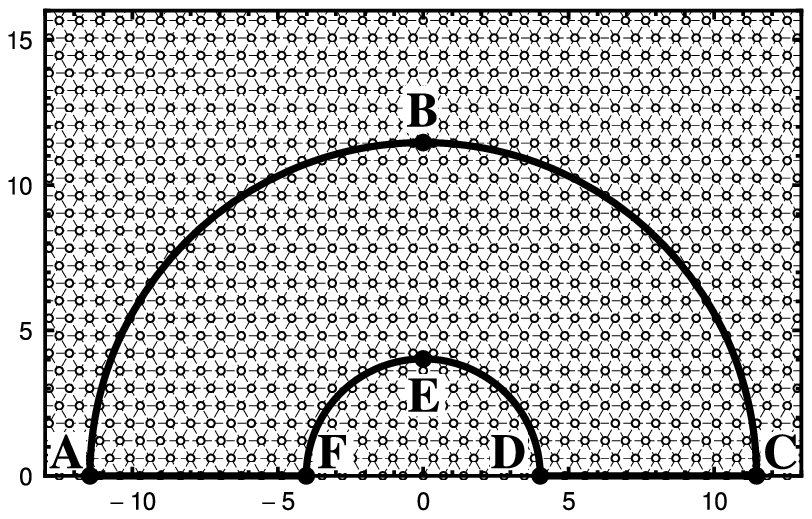}

\includegraphics{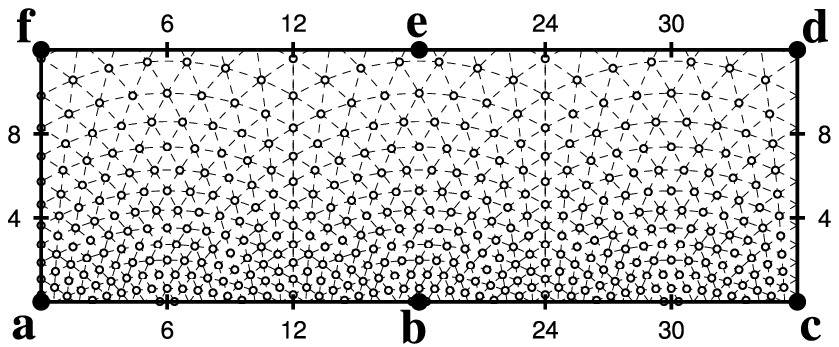}

\caption{ A conformal transformation 
is applied to the semiannular section of a regular hexagonal lattice shown in (a) 
to create the conformal crystal structure shown in (b)\cite{Suppl}. 
Points $A-F$ in (a) are mapped by the transformation to points $a-f$ in 
(b) respectively. 
The straight contour lines connecting nearest neighbor
lattice points in (a) are bent into arcs in (b), 
but the local six-fold ordering of the 
lattice points
is maintained. 
Pinning sites are placed at the vertices formed by the 
intersections of the contour lines in (b). 
 }
\label{fig:0} 
\end{figure}

\begin{figure}
\includegraphics{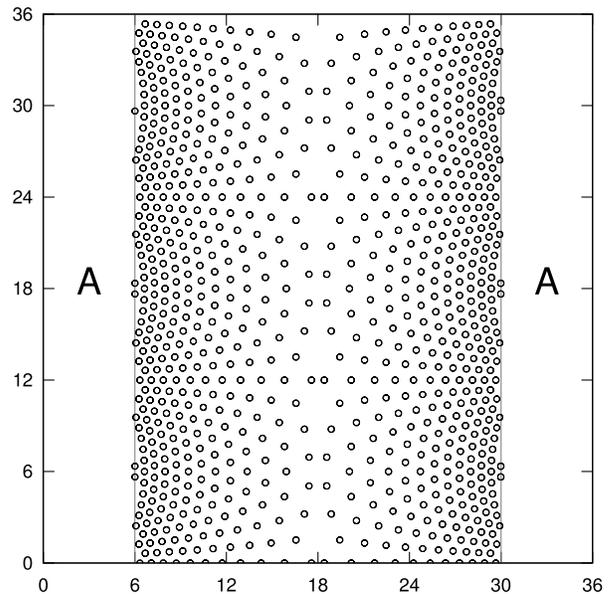}
\caption{ The gradient-driven
sample geometry consists of two conformal crystals facing each other
in the pinned region. Open circles are pinning site locations.
Vortices are added to or subtracted from the
pin-free region labeled ``A'' on the left and right sides of the image. }
\label{fig:1}
\end{figure}

To examine vortex pinning by conformal crystalline substrates, 
we numerically compute the magnetization $M$, which is a measure of
the pinning strength. We compare the effectiveness of the conformal
pinning to that of random and periodic pinning arrays with the same
number of pinning sites. The conformal array produces significantly
stronger pinning over a much larger range of fields than these other
arrays, with the exception that the pinning by periodic arrays
is strongest 
only in the vicinity of the matching fields \cite{Periodic}. 
By comparing our results
to those obtained for random pinning with an equivalent gradient,
we demonstrate that it is not simply the non-uniform density of the
conformal array that gives rise to the pinning enhancement; rather,
it is the preservation of the hexagonal ordering of the pinning by
the conformal map that plays the most critical role. 
Another feature that gives the conformal pinning
an advantage over periodic pinning for incommensurate fields is that
even though local hexagonal ordering is present, the overall arc-like arrangement 
of the pinning sites (apparent in Figs.~\ref{fig:0}(b) and \ref{fig:1}) 
prevents the formation of straight-line channels along which vortices 
can easily flow. 
In addition
to vortex pinning in type-II superconductors, the effects of conformal
crystalline substrates on ordering or dynamics of a monolayer of particles
could also be studied for vortices in Bose-Einstein condensates on
optical lattices \cite{Tung} or colloidal particles on optically
created substrate arrays \cite{Grier}. The enhanced pinning also
suggests that conformal arrays could be used to increase friction
for particle-surface interactions.

Conformal crystals are a class of two-dimensional
(2D) structures created by the application of a conformal 
(angle-preserving) transformation
to a regular lattice in the complex plane \cite{Rothen,C}. Figure~\ref{fig:0}
illustrates a conformal crystal obtained via the transformation of
a hexagonal lattice. The contour lines 
connecting nearest neighbors, which are straight 
lines for the original hexagonal lattice, are bent into arcs 
but still cross at angles of $\pi/3$, preserving the sixfold coordination of
individual pinning sites in spite of the clear density gradient.
To create a pinning lattice, we place
pinning sites at the vertex locations where the contour lines intersect.
Conformal crystal structures have been studied experimentally for
repulsively interacting magnetic spheres confined to a 2D container
that is tilted so that the gravitational force on the particles produces
a mechanically stable but nonuniform crystal \cite{C}. Other systems
where conformal crystals arise include foams under an external field
\cite{Weaire,Ma} and large arrays of classical Coulomb charges in
confined circular potentials where local triangular ordering occurs
along curved lattice lines \cite{Moore}.

{\it Simulation--}
We conduct a flux gradient density simulation 
of the type previously used to study vortex critical states and magnetization
with random \cite{N,M} and periodic pinning arrays \cite{Periodic}.
Figure~\ref{fig:1} shows the
outer
pin-free region surrounding a central pinned region that consists
of two conformal crystals placed with their highest density regions
adjacent to the pin-free region. Details on the construction of the conformal crystal are given in the supplemental material\cite{Suppl}.
We use periodic boundary conditions
in the $x$ and $y$-directions and 
consider a $36\lambda\times36\lambda$
system with pinned region extending from $x=6\lambda$ to $30\lambda$, where
$\lambda$ is the
penetration depth. 
This geometry was previously shown to be large enough to capture accurately
the behavior of the magnetization curves \cite{N,M,Periodic}. 

The dynamics of vortex $i$ are obtained by integrating the 
overdamped equation 
$\eta(d{\bf R}_{i}/dt)={\bf F}_{i}^{vv}+{\bf F}_{i}^{vp}.$
$\eta$ is the damping constant which is set equal to unity. The vortex-vortex
interaction force is 
${\bf F}_{i}^{vv}=\sum_{i=1}^{N_{v}}s_is_jF_{0}K_{1}(R_{ij}/\lambda){\hat{{\bf R}}}_{ij}$,
where $K_{1}$ is the modified Bessel function, 
${\bf R}_{i}$ is the location of vortex $i$,
$R_{ij}=|{\bf R}_i-{\bf R}_j|$,
${\hat{{\bf R}}}_{ij}=({\bf R}_{i}-{\bf R}_{j})/R_{ij},$
$F_{0}=\phi_{0}^{2}\pi\mu_{0}\lambda^{3}$, and $\phi_{0}$ is the
elementary flux quantum. 
The sign prefactor $s_i$ is $+1$ for a vortex and $-1$ for an antivortex.
The pinning sites are modeled as $N_{p}$
non-overlapping parabolic traps with 
${\bf F}_{i}^{vp}=\sum_{k=1}^{N_{p}}(F_pR_{ik}^p/R_{p})
\Theta((R_{p}-R_{ik}^p)/\lambda){\hat{{\bf R}}}_{ik}^p$,
where $R_k^p$ is the location of pinning site $k$, 
$R_{ik}^p=|{\bf R}_i-{\bf R}_k^p|$, 
$\hat{\bf R}_{ik}^p=({\bf R}_i-{\bf R}_k^p)/R_{ik}^p$, 
$\Theta$ is the Heaviside step function, 
$R_{p}$ is the pinning radius that we fix to $R_{p}=0.12\lambda$, and 
$F_p$ is an adjustable parameter controlling the strength of the pinning force.
All forces are measured in units of $F_{0}$ and all lengths in units
of $\lambda$. The external field is measured in units of $H_{\phi}$,
the field at which the average unit density of vortices equals
the average unit density of pinning sites. 

To perform a complete field sweep, we begin with zero vortex density
and then quasistatically add vortices in the unpinned region 
(labeled ``A'' in Fig.~\ref{fig:1}) 
at randomly chosen nonoverlapping positions
until the desired maximum external field value is reached. 
As the vortex or antivortex density builds
up in the pin-free region, the vortices or antivortices 
drive themselves into the
pinned region due to their own repulsive interactions, creating a
gradient in the flux density within the pinned region \cite{N,M,Periodic}.
We then remove vortices from the pin-free region until
the vortex density in this region reaches zero. 
To reverse the field, we add antivortices, which repel each other
but are attracted to vortices, to the unpinned region.
When a vortex
and antivortex come within a small distance 
($0.3\lambda$) of each other, they are
both removed from the system to simulate an annihilation event. To complete an entire magnetization loop, we continue to add anti-vortices until the external field reaches its most negative value, and then remove antivortices from
the pin-free region 
to bring the external field back up to zero. The
average magnetization $M$ is calculated as the difference between the
flux density $H$ in the unpinned region and the density $B$ in
the pinned region, $M=-(1/4\pi V)\int(H-B)dV$, where $V$ is the
sample area. The critical current $J_{c}$ is proportional to the
magnetization.

\begin{figure}
\includegraphics[width=3.5in]{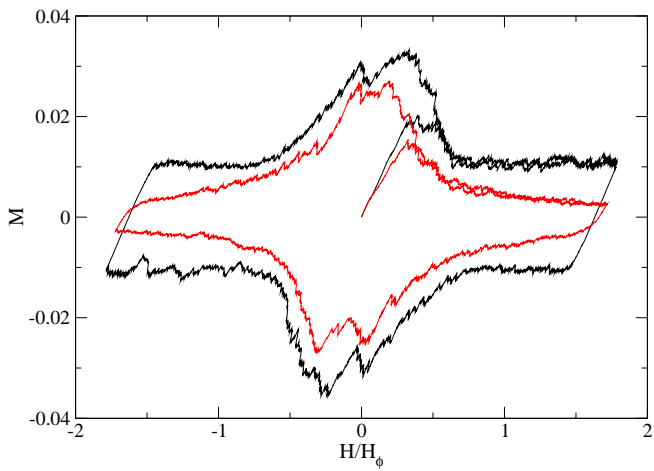} 
\caption{ The magnetization $M$ vs $H/H_{\phi}$, where $H_{\phi}$ is
the matching field at which the vortex density equals the average pinning
density, in samples with $n_p=1.0$ and $F_p=0.55$. 
Outer dark (black) curve: a sample with a conformal pinning
array (CPA). 
Inner light (red)
curve: a sample with a uniformly dense random arrangement of pinning
sites. 
The CPA produces a much
higher value of $M$ at all fields. 
}
\label{fig:3}
\end{figure}

\begin{figure}
\includegraphics[width=3.5in]{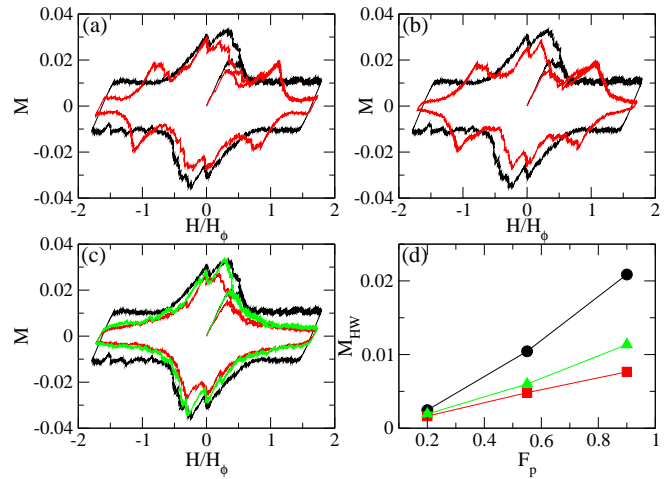} 
\caption{ 
(a-c) $M$ vs $H/H_{\phi}$ plots for samples with $F_{p}=0.55$
and average pin density $n_{p}=1.0$. (a) Dark (black) curve: the
CPA system from Fig.~\ref{fig:3}; 
light (red) curve: a uniform square pinning array. 
(b) Dark (black) curve: the CPA system; 
light (red) curve: a uniform triangular pinning array. 
$M$ is higher for both the uniform periodic pinning systems than
for the CPA at the matching field near $H/H_{\phi}=1.0$, but the
CPA has the highest value of $M$ at all other fields. 
(c) Outer dark
(black) curve: the CPA system;
middle light (green) curve: a random arrangement of pinning sites
with an equivalent pinning gradient; 
inner light (red) curve: the uniform random array. 
Here the gradient in the random pinning increases $M$ compared to
the uniform random pinning by only a small amount for most fields.
(d) $M_{HW}$, the half-width of the magnetization loop at $H/H_{\phi}=1.0$, 
for the CPA (circles), random array with CPA-equivalent pinning gradient 
(triangles), and uniform random pinning array (squares) 
for varied $F_{p}$. There is a consistent enhancement of $M$ in the CPA. }
\label{fig:vt} 
\end{figure}

{\it Results--}
In Fig.~\ref{fig:3} we plot the complete hysteresis loop $M$ vs $H/H_{\phi}$
for the conformal pinning array (CPA) and the uniformly dense random
pinning array. Each sample contains the same number $N_{p}$ of pinning
sites with an average pinning density of $n_{p}=1.0$ and with $F_{p}=0.55$.
We find that $M$ is much higher at all fields for the CPA than for
the random pinning, and that at its highest point, $M$ for the CPA
is almost four times higher than for the random pinning. Although
the CPA has local triangular ordering, we observe no peaks or other
anomalies in $M$ at integer matching or fractional matching multiples
of $H/H_{\phi}$ of the type found for uniformly dense periodic pinning
arrays \cite{Periodic}. 
The flux profiles plotted in Suppl.~Fig.~1\cite{Suppl} show that the random
array produces a Bean-like profile that becomes shallower as $H$
increases. In contrast, at higher fields the CPA does not have a uniform
flux gradient but instead develops a double slope profile, with a
larger flux gradient near the edge of the sample and a much shallower
or nearly flat flux profile in the center of the sample. As $H$ increases,
the sharper slope region decreases in width and is replaced by the
shallow slope region.
We have examined a range of pinning forces
and find that the magnetization for the CPA is consistently
higher than for random pinning arrays, as shown in 
Fig.~\ref{fig:vt}(d)
where we plot $M_{HW}$, the half-width of the magnetization loop
at $H/H_{\phi}=1.0$, at varied $F_{p}$ for these
array types. 

We next address whether the conformal pinning arrays produce higher
pinning compared to other previously studied types of non-random pinning
arrays. In Fig.~\ref{fig:vt}(a) we plot $M$ vs $H/H_{\phi}$ for the CPA and
a square pinning array with the same pinning density and strength,
and in Fig.~\ref{fig:vt}(b) 
we plot the same quantity for the CPA and a triangular
pinning array. In both cases, $M$ for the CPA is higher over most
of the range of $H/H_{\phi}$ except at the first matching field,
where the periodic pinning arrays produce a higher value of $M$ due
to a commensurability effect. This shows that although a periodic
pin structure can strongly enhance the pinning, this enhancement occurs
only for a very specific matching field. In contrast, the CPA produces
a significant enhancement of the pinning over a very broad range of
fields, extending well above the first matching field.

Since the CPA has a pinning gradient, it could be possible that any
type of pinning array with an equivalent gradient would also exhibit
a pinning enhancement compared to uniform pinning arrays and could
be just as efficient at pinning as the CPA. We find that this is not
the case. In Fig.~\ref{fig:vt}(c) we plot $M$ versus $H/H_{\phi}$ for the
CPA and a random pinning array with an equivalent pinning gradient
and the same average pinning density. For comparison we also show
the uniform random pinning array. The random pinning with a gradient exhibits
a modest enhancement of $M$ compared to the uniform random pinning
array; however, both arrays give $M$ values that are significantly
smaller than the CPA for all but the very lowest fields. This result
indicates that it is the other properties of the CPA such as the
preservation of local sixfold ordering, and not merely the pinning
gradient, that are responsible for the enhanced pinning. The arching
structure of the CPA suppresses certain modes of vortex motion. For
example, for random pinning arrays it has been found that certain
regions where the pinning density is slightly lower lead to the formation
of persistent river-like flow patterns \cite{M}; these structures
are unable to form in the CPA. For periodic pinning arrangements,
easy vortex flow occurs along the symmetry directions of the pinning
array \cite{Sih,I}, causing a drop in the critical current above
a commensurate field.
Such motion is suppressed in the CPA due to the
arching structure.

\begin{figure}
\includegraphics[width=3.5in]{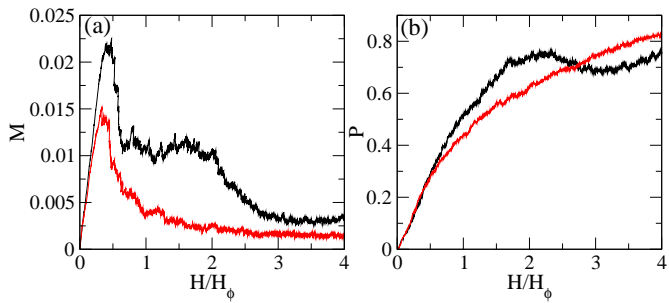} 
\caption{ (a) The high field behavior of $M$ vs $H/H_{\phi}$ for samples
with $F_{p}=0.55$ and $n_{p}=1.0$ on the initial ramp up only. Dark
(black) curve: CPA; light (red) curve: uniform random array. Although
$M$ is higher for the CPA than for the random array over the entire
range of fields, there is a drop in $M$ for $H/H_{\phi}>2.0$ in
the CPA. (b) The fraction of occupied pinning sites 
$P$ vs $H/H_{\phi}$
for the CPA (dark black curve) and uniform random pinning (light red
curve). 
$P$
is initially higher for the CPA but drops at $H/H_{\phi}=2$,
coinciding with the drop in $M$ in (a), while 
$P$ for the
random pinning sample increases monotonically with increasing field. }

\label{fig:4} 
\end{figure}

In Fig.~\ref{fig:4}(a) we compare $M$ versus $H/H_{\phi}$ for the CPA and
uniform random arrays in the first quarter of the magnetization loop
over a much larger range of fields up to $H/H_{\phi}=4.0$, and in
Fig.~\ref{fig:4}(b) we plot the corresponding pin occupancy 
$P$, which is
the fraction of pinning sites occupied by vortices. The enhanced pinning
for the CPA is the most pronounced below $H/H_{\phi}=2.0$. The enhancement
decreases above this field but remains larger than the random pinning
array for all fields. 
For the random pinning array, 
$P$ monotonically increases over the
entire range of $H/H_{\phi}$. In contrast, after running well above
the 
$P$ value for the random pinning array at lower fields, 
$P$
for the CPA rolls over and begins to decrease with increasing field
above $H/H_{\phi}\approx2$, correlated with the decrease in $M$.
This is also the field at which the higher gradient region 
seen in
Suppl.~Fig.~1(a)
disappears from the sample \cite{Suppl}. Just below this field, all
of the pinning sites near the edge of the sample are occupied, and
since these pinning sites are the most closely spaced, as additional
vortices enter the sample, the vortex-vortex interactions become strong
enough to push some of the vortices out of the pins, leading to the
drop in 
$P$ and $M$. For the random pinning array, there are always
some empty pinning sites near the edge of the sample in places where
two pins happen to be in close proximity, so that the vortex-vortex
interaction energy would be prohibitively high if both pins were occupied
simultaneously. As the field increases, these pinning sites gradually
become occupied. Even though 
$P$ for the CPA falls below 
$P$ for
the random array at higher fields, the pinning enhancement remains
significantly stronger for the CPA. 

In summary, we 
demonstrate strongly enhanced vortex pinning by a conformal crystal array
of pinning sites. The conformal crystal is constructed by a conformal
transformation of a hexagonal lattice, producing a nonuniform structure
with a gradient where the local sixfold coordination of the pinning
sites is preserved, and with an arching effect. The conformal pinning
arrays produce significantly enhanced pinning over a much wider range
of field than that found for other pinning geometries with an equivalent
number of pinning sites, such as random, square, and triangular. We
show that the pinning enhancement is not simply due to the pin density
gradient, but is also due to the preservation of the sixfold coordination
of the pinning sites and to the arching effects that prevent the formation
of easy channels of vortex flow. 
The conformal pinning arrays do not show
any pronounced enhancement at the matching fields of the type found
for periodic pinning arrays; however, there is a drop in the magnetization
at higher fields that is correlated with a drop in the occupancy of
the pinning sites.  Our results will be important for applications where
high critical currents are required, and
can be generalized to stabilizing or
pinning other systems, such as colloidal particles, 
that do not form uniform crystalline structures.

This work was carried out under the auspices of the NNSA of the U.S.
DoE at LANL under Contract No. DE-AC52-06NA25396.


\section*{SUPPLEMENTARY MATERIAL}
\subsection*{for ``Strongly Enhanced Vortex Pinning by Conformal Crystal Arrays''}


\setcounter{figure}{0}
\begin{figure}[b]
\includegraphics[width=3.5in]{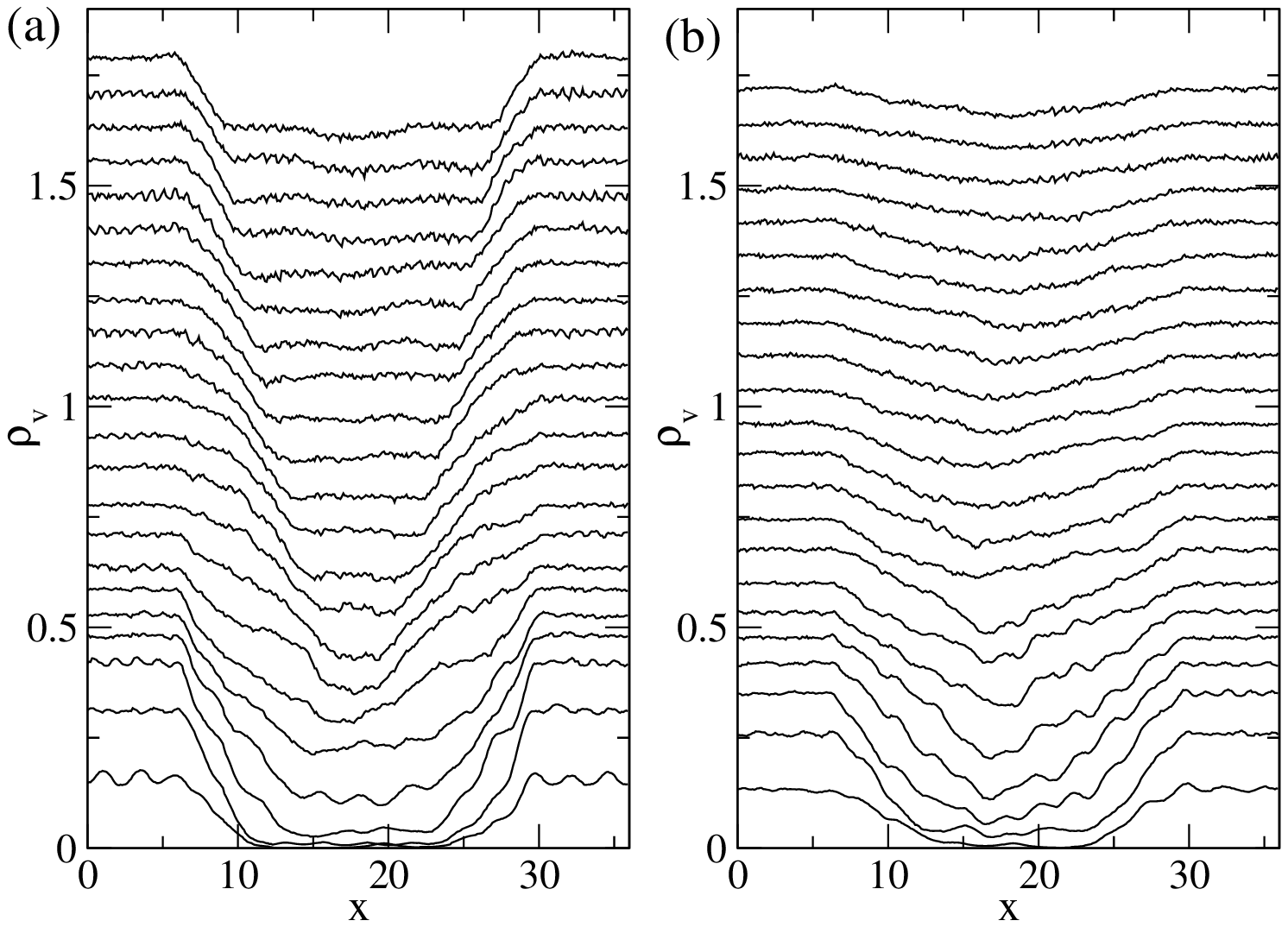}
\caption{Data from the samples in Fig.~3 of the main text with
$n_p=1.0$ and $F_p=0.55$.
(a) The average flux density $\rho_{v}$ across
the sample plotted for different times during the ramp up portion
of the hysteresis curve 
for the conformal pinning array (CPA). (b) $\rho_{v}$ for the
random pinning array.
The random array exhibits a Bean profile while
the CPA has two regions inside the sample: a higher flux gradient
at the edges of the pinned region and a small gradient in the middle. 
}
\end{figure}

\vskip2pc

\section{Construction of the Conformal Crystal}
To obtain the conformal crystal structure, we first situate a regular hexagonal lattice in the complex plane: that is, we encode the location $\left(x,y\right)$ of each lattice site into a complex number $z=x+iy$. With lattice constant $b$, the lattice sites will be located at 
\begin{equation}
z=n_1\cdot\left(1\cdot b\right) + n_2\cdot\left(e^{i\pi /3}\cdot b\right) 
\end{equation}
where $n_1$ and $n_2$ are arbitrary integers.

We then consider a semiannular region of this lattice as shown in 
Fig.~1(a) of the main text,
defined by $\text{Im} z\geq 0$ and $r_\text{in}\leq \left|z\right| \leq r_\text{out}$, and apply a conformal (angle-preserving) transformation mapping $z$ to $w\equiv u+iv$, given by 
\begin{equation}
w=\frac{\pi}{2\alpha}+\frac{1}{i\alpha}\ln\left(i\alpha z\right)
\end{equation}
where $\alpha\equiv\frac{1}{r_\text{out}}$. The semiannular region in the $z$-plane shown in 
Fig.~1(a) of the main text will be mapped to the 
rectangular region in the $w$-plane shown in 
Fig.~1(b) of the main text,
specified by 
\begin{equation*}
0\leq u \leq\pi r_\text{out}\quad\text{and}\quad 0\leq v \leq r_\text{out}\ln\left(\frac{r_\text{out}}{r_\text{in}}\right).
\end{equation*}

\subsection{Explicit Construction}
The structure of the final conformal crystal is completely determined by specifying the three constants $b$, $r_\text{out}$, and $r_\text{in}$. The locations of the sites of the conformal crystal can be generated directly by the expressions  
\begin{subequations}
 \begin{align}
u & =  r_\text{out} \cdot \left\{ \frac{\pi}{2}-\tan^{-1}\left(\frac{2n_{1}+n_2}{n_{2}\sqrt{3}}\right) \right\}  \\
v & =  r_\text{out}\ln\frac{r_\text{out}}{b\sqrt{n_{1}^{2}+n_{2}^{2}+n_{1}n_{2}}} 
 \end{align}
\end{subequations}
where $\left(n_1,n_2\right)$ range over all pairs of integers satisfying the two constraints, $n_2\geq 0$ and 
\begin{equation*}
r_\text{in}^{2}\leq b^{2}\left(n_{1}^{2}+n_{2}^{2}+n_{1}n_{2}\right)\leq r_\text{out}^{2}.
\end{equation*}
To obtain a conformal crystal of length $u_\text{max}$ and width $v_\text{max}$, one chooses $r_\text{out}$ and $r_\text{in}$ to have the values 
\begin{equation*}
r_\text{out}=\frac{u_\text{max}}{\pi},\ r_\text{in}=\frac{u_\text{max}}{\pi}\exp\left(-\frac{\pi v_\text{max}}{u_\text{max}}\right).
\end{equation*}
$b$ can be chosen to obtain a desired final density $\rho$ of lattice sites:
\begin{equation*}
b^2=\frac{1-\left(r_\text{in}/r_\text{out}\right)^2}{\rho\sqrt{3}\cdot\ln\left(r_\text{out}/r_\text{in}\right)}
\end{equation*}
Alternatively, it can be set equal to a desired minimum distance between pinning sites in the final structure. (Note that in 
Fig.~1 of the main text, the lengths of arc $ABC$ and line $abc$ are equal.) 

Figure~1 of the main text was obtained using $r_\text{out}=36/\pi$, $r_\text{in}=r_\text{out}e^{-\pi/3}$, and $b^2=\left(1-\exp\left(-2\pi/3\right)\right)\cdot\left(\sqrt{3}/\pi\right)$. These parameter values were calculated from $u_\text{max}=36$, $v_\text{max}=12$, and $\rho=1$.

\section{Flux density profiles}
In Supplemental Fig.~1(a) we plot the flux density profiles obtained by
averaging the flux density in the $y$ direction for the first ramp up of
the field in the hysteresis loop for the conformal pinning array (CPA)
illustrated in Fig.~3 of the main text, and in Supplemental Fig.~1(b) we show
the flux density profiles for the uniform random pinning array from Fig.~3 of
the main text.
The random
array produces a Bean-like profile that becomes shallower as $H$
increases. In contrast, at higher fields the CPA does not have a uniform
flux gradient but instead develops a double slope profile, with a
larger flux gradient near the edge of the sample and a much shallower
or nearly flat flux profile in the center of the sample. As $H$ increases,
the sharper slope region decreases in width and is replaced by the
shallow slope region.  


\end{document}